\documentstyle [11pt]{article}

\newcommand{\beq}{\begin{equation}}
\newcommand{\eq}{\end{equation}}
\newcommand{\bega}{\begin{eqnarray}}
\newcommand{\ega}{\end{eqnarray}}

\begin{document}

\title{A combined mathematica--fortran program package for
analytical calculation of
the matrix elements of the microscopic cluster model}

\author{K\'alm\'an Varga
\\
Institute  of Nuclear Research of the Hungarian Academy of
Sciences
\\
(MTA ATOMKI) Debrecen, Hungary}
\maketitle
\begin{abstract}
We present a computer code that analytically
evaluates the matrix elements
of the microscopic nuclear Hamiltonian and unity operator
between Slater
determinants of displaced gaussian single particle orbits.
Such matrix elements appear in the generator coordinate
model and the resonating group model versions of the
microscopic multicluster calculations.
\end{abstract}

\newpage
{\par\noindent\Large \bf PROGRAM SUMMARY}
8pt

{\par\noindent\small {\sl Title of the program:} MCKER}
\vskip 8pt
{\par\noindent\small {\sl Catalog number:}}
\vskip 8pt
{\par\noindent\small {\sl Program obtainable from:} CPC Program
Library, Queen's University of Belfast, N. Ireland (see application form
in this issue)}
\vskip 8pt
{\par\noindent\small {\sl Licensing provisions:} None}
\vskip 8pt
{\par\noindent\small {\sl Computer:} IBM PC 386DX}
\vskip 8pt
{\par\noindent\small {\sl Operating system:} DOS 3.2}
\vskip 8pt
{\par\noindent\small {\sl Programming language used:} Mathematica v.1.2,
Fortran}
\vskip 8pt
{\par\noindent\small {\sl No. of lines in distributed program,including
test data etc.:}}
\vskip 8pt
{\par\noindent\small {\sl Nature of physical problem}
\par\noindent
This computer code
contructs matrix elements to be used in microscopic studies of light
multicluster system.}
\vskip 8pt
{\par\noindent\small {\sl Method of solution}
\par\noindent
We analytically evaluate the determinants and inverses appearing
in the matrix elements of Slater determinants of
nonorthogonal single particle states,
using the Mathematica symbolic manipulation language. A fortran code
is used to substitute the space part of the single particle
matrix elements.}
\vskip 8pt
{\par\noindent\small {\sl Restrictions on the complexity of the problem}
\par\noindent
Varies with available RAM and type of computer. Use of the available RAM
depends on the number of clusters and nucleons.
\vskip 8pt
{\par\noindent\small {\sl Typical running time}
\par\noindent
For the example given in the test run input, approximately 300 s.

\newpage

{\par\noindent\Large \bf LONG WRITE-UP}

\vskip 48pt
\section{Introduction}

The microscopic nuclear cluster model provides an unquestionably
successful description of light nuclear systems \cite{Tang},
\cite{Langanke}.
The main difficulty in performing microscopic cluster-model
calculation is the evaluation of the microscopic Hamiltonian
between Slater determinants of nonorthogonal single particle-states.
This calculation becomes especially tedious for multicluster systems.
The need for analytical form of these matrix elements calls for
application of symbolic manipulation languages.

In this paper the single particle states of the
Slater determinants are displaced gaussian functions, that is,
we determine the matrix elements of the ``generator coordinate model''
(GCM) \cite{Tang}. To achieve this, we combine the advantage of  computer
languages the Fortran
and the Mathematica \cite{Wolfram}: With the help of Mathematica
we can  derive analytical expressions for the matrix elements,
with Fortran we will sort out and specialize the formulae.

Our program is applicable for a general N-cluster system of $0s$
clusters of common oscillator width. The generalization of the
program for nonequal width parameters or for higher harmonic
oscillator orbits is straightforward. We hope that our program
brings the technique of the nuclear cluster model in the reach of nuclear
physicist.

\section{Formulation of the problem}

Let us consider an $N$-cluster system, where each cluster consists
of $A_i \  (i=1,...,N, \  \sum_{i=1}^N A_i = A) $ nucleons, and the $k$th
nucleon occupies the single particle state $\psi_{j}(x_k)$ (see
Figure 1). The
notation $x_k$ stands for the space-, spin- and
isospin-coordinate of the $k$th nucleon:
$x_{k}=({\bf r}_{k},\sigma_{k},\tau_{k})$.
We assume that  the nucleons of the
$i$th cluster occupy the single particle states of the form
\beq
\psi_{j}(x_{k})=\varphi_{i}({\bf r}_{k}) \chi_{\sigma_j}(k)
\chi_{\tau_j}(k) , \ \ \ \ (j=A_{i-1},...,A_{i}),
\label{eq:1}
\eq
where $\varphi_i({\bf r}_i)$, $\chi_{\sigma_j}(k)$ and
$\chi_{\tau_j}(j)$ are the space, spin and  isospin parts of the
wavefunction, respectively. (Note the cluster label $i$ on
$\varphi_i({\bf r}_i)$.)
We restrict ourselves to the case, where the space and the
isospin parts of the single-particle function of the nucleons of
the $A$ particle system are fixed, but the spins may have
different orientation. A given set of spin quantum numbers is
labelled by $\alpha$.

The spin and the isospin functions of different states are orthogonal
\beq
\langle
\chi_{\sigma_k} \vert \chi_{\sigma_l} \rangle =\delta_{\sigma_k
\sigma_l} ,
\label{eq:2}
\eq
\beq
\langle
\chi_{\tau_k} \vert \chi_{\tau_l} \rangle =\delta_{\tau_k \tau_l} ,
\label{eq:3}
\eq
the space-part of single-particle states overlap, defining an
overlap matrix
\beq
b(i,j)=\langle \varphi_i \vert \varphi_j \rangle .
\label{eq:4}
\eq

The wave function of the A-particle system is given by linear combination
\beq
\Psi=\sum_{\alpha} a_{\alpha} \Psi^{\alpha}
\eq
of the Slater determinants
\beq
\Psi^{\alpha}={1 \over \sqrt{A!}} {\rm det}
\lbrace \psi_{i}^{\alpha}(x_{j}) \rbrace
\label{eq:5}
\eq
of the different sets of single particle
states.

The matrix elements of Slater determinants can be calculated
using well-known rules \cite{Brink}, \cite{Horiuchi}. In the
following we give a concise summary of these formulae to
establish the formalism of our program.

The overlap of two Slater determinants can be written as
\beq
\langle \Psi^{\alpha} \vert \Psi^{\beta} \rangle =
{\rm det} \lbrace \langle \psi_{k}^{\alpha}
\vert \psi_{l}^{\beta}\rangle \rbrace  .
\label{eq:6}
\eq
By using eqs. (\ref{eq:2}), (\ref{eq:3}) and (\ref{eq:4})
the overlap of the single-particle states becomes
\beq
B_{kl}^{\alpha \beta}=\lbrace \langle \psi_{k}^{\alpha}
\vert \psi_{l}^{\beta}\rangle \rbrace =
b(i,j) \delta_{\sigma_k^\alpha \sigma_l^{\beta}}
\delta_{\tau_k \tau_l} , \ \ \ \ (k=A_{i-1},...,A_i , \ \ \ \
l=A_{j-1},...,A_j) , \ \ \ (i,j=1,...,N)
\label{eq:7}
\eq
thus to determine the overlap we must calculate the determinant
of the matrix
\beq
\label{eq:8}
B^{\alpha \beta}=\lbrace B_{ij}^{\alpha \beta}
\rbrace \ \ \ \ (i,j=1,...,A) .
\eq

The matrix elements of the one-body operator are given by
\beq
\label{eq:9}
\sum_{i=1}^{A} \langle \Psi^{\alpha} \vert {\cal O}_i^{(1)}
\vert \Psi^{\beta} \rangle=
\sum_{i=1}^{A} \sum_{j=1}^A \langle \psi_{i}^{\alpha}
\vert {\cal O}^{(1)} \vert
\psi_{j}^{\beta} \rangle  \Delta_{ij}^{\alpha \beta}  ,
\eq
where $\Delta_{ij}^{\alpha \beta}$ is the cofactor of $B^{\alpha
\beta}$ (i.e. the subdeterminant,
obtained by crossing out the $i$th row and the $j$th column of
$B^{\alpha \beta}$
and multiplying it with the phase $(-1)^{i+j}$).

The matrix elements of the two-body operator can be determined
from the formula
\beq
\label{eq:10}
\sum_{i < j}^{A} \langle \Psi^{\alpha} \vert {\cal O}_{ij}^{(2)} \vert
\Psi^{\beta} \rangle=
\sum_{{i < j} \atop {k < l}}^{A}
\langle \psi_{i}^{\alpha} \psi_{j}^{\alpha}
\vert {\cal O}^{(2)} \vert \psi_{k}^{\beta} \psi_{l}^{\beta}
\rangle_a (-1)^{i+j+k+l} {\rm det} D_{ijkl}^{\alpha \beta}  ,
\eq
where
\beq
\vert \psi_k \psi_l \rangle_a =
\vert \psi_k \psi_l \rangle -
\vert \psi_l \psi_k \rangle  ,
\nonumber
\eq
and where $D_{ijkl}^{\alpha \beta}$ is
an $(A-2)\times (A-2)$ matrix built up
of elements of $B^{\alpha \beta}$ by leaving
out its rows $i$ and $k$ and its
columns $j$ and $l$.

In this paper we limit ourselves to the case of spin-isospin
independent one-body operators, and for the two-body operator of
the form
\beq
\label{eq:11}
{\cal O}_{ij}^{(2)}=V(i,j)(w + m {P_{ij}}^r +
b {P_{ij}}^{\sigma} - h {P_{ij}}^{\tau})
\eq
where $P_{ij}$ are the space- , spin- , and isospin-exchange operators.
These restrictions are not too serious, one can easily
generalize the program to allow for other type of operators.

\section{Mathematica part}

The analytical calculations of the determinants and sums of eqs.
(\ref{eq:6}), (\ref{eq:9}) and (\ref{eq:10}) can be easily
performed by using the algebraic computer language Mathematica.

By the aid of Mathematica we first evaluate the single particle
overlaps (see eqs. (\ref{eq:2}) and (\ref{eq:3}) ) and then
build up the matrices $B$, $\Delta^{ij}$ and $D^{ijkl}$ and
carry out the operations in eqs. (\ref{eq:6}), (\ref{eq:9}) and
(\ref{eq:10}) . The matrix elements (``kernels'') have the form

\par\noindent
-- overlap :
\beq
\label{eq:12}
\langle \Psi^{\alpha} \vert \Psi^{\beta} \rangle =
\sum_{n=1}^{n_o} c_n^{o} \prod_{l_n=1}^{m_n}
b(i_{l_n},j_{l_n})^{k_{l_n}} ,
\eq
\par\noindent
-- one-body :
\beq
\label{eq:13}
\sum_{i=1}^{A} \langle \Psi^{\alpha} \vert {\cal O}_i^{(1)}
\vert \Psi^{\beta} \rangle=
\sum_{n=1}^{n_{ob}} c_n^{ob} t(p_n,q_n)
 \prod_{l_n=1}^{m_n} b(i_{l_n},j_{l_n})^{k_{l_n}} ,
\eq
where $t(p,q)=\langle \varphi_p \vert {\cal O}^{(1)} \vert
\varphi_q \rangle$,
\par\noindent
-- two-body :
\beq
\label{eq:14}
\sum_{i < j}^{A} \langle \Psi^{\alpha} \vert {\cal O}_{ij}^{(2)}
\vert \Psi^{\beta}
\rangle=
\sum_{\nu=1}^{\nu_{tb}}  (w {\sl W_\nu}+ m {\sl M_\nu}+ b {\sl B_\nu}+h
{\sl H_\nu}) v(p_\nu,q_\nu;r_\nu,s_\nu)
\sum_{n=1}^{n^{tb}_\nu} c_{n}^{tb}
\prod_{l_n=1}^{m_n} b(i_{l_n},j_{l_n})^{k_{l_n}} ,
\eq
where $
\langle \varphi_{p_\nu} \varphi_{q_\nu} \vert {\cal O}^{(2)} \vert
\varphi_{r_\nu} \varphi_{s_\nu} \rangle
=v(p_\nu,q_\nu;r_\nu,s_\nu)(w{\sl W_\nu}+m{\sl M_\nu}+b{\sl
B_\nu}+h{\sl H_\nu})$.
All the coefficients, indices, limits of summations and products
are to be determined by Mathematica.
A detailed example of these formulae can be found in section 3.3.

\subsection{Input data}
The input data are to be read from the input file ``{\sl nuc.dat}''.
Detailed input data specification:
\par\noindent 1) NumClus: Number of clusters.
\par\noindent 2) NumPar: Number of particles in the clusters.
(Numpar(i), i=1,...,Numclus)
\par\noindent 3) isospin: Charge number of the particles, 1 stands for
protons and 0 stands for neutrons,
(isospin(i), i=1,...,A).
\par\noindent 4) SpinConf: Number of spin configurations.
\par\noindent 5) spin: Spin of the particles,
1 stands for spin up, 0 stands for spin down,
(spin(i,j), i=1,...,A, j=1,...,SpinConf).
\par\noindent 6) $a_\alpha$: Clebsch-Gordan coefficients
of the spin configurations,
($\alpha$=1,...,SpinConf).

\subsection{Output data}
To calculate the overlap, the one-body and the two-body kernels,
one should use the programs ``{\sl ok.m}'', ``{\sl obk.m}'' and
``{\sl tbk.m}'',
respectively. The output is written into the files ``{\sl ok.out}'',
``{\sl obk.out}'' and ``{\sl tbk.out}'', accordingly.
As output we get the coefficients and indices of eqs. (\ref{eq:12}),
(\ref{eq:13}) and (\ref{eq:14}) ordered according to Table 1.
In the output files the results for different spin
configurations follow in consecutive order.

\subsection{Test run}
As a test run, we consider the example of $^6$He=$\alpha+n+n$.
This is a three-cluster system, the two outer neutrons
may have two different spin configurations with antiparallel spins
(see the test run input).
The combination coefficients are
$a_1=-a_2=<{1\over 2} {1\over 2}{1\over 2}  -{1\over 2} \vert 0 0 >
= 1/\sqrt{2} $.
The package must be loaded while running the Mathematica program.
The program should work with any Mathematica implementation on any
machine without modifications.
The output of ``{\sl ok.m}'' (ok.out) corresponds to the overlap
\bega
\langle \Psi \vert \Psi \rangle & = &
b(1,1)^2 b(1,2) b(1,3) b(2,1) b(3,1) - b(1,1)^3 b(1,3) b(2,2)
b(3,1) - b(1,1)^3 b(1,2) b(2,1) b(3,3)
\nonumber
\\
& + &
b(1,1)^4 b(2,2) b(3,3) -  b(1,1)^2 b(1,2) b(1,3) b(2,1) b(3,1)
+ b(1,1)^3 b(1,2) b(2,3) b(3,1)
\nonumber
\\
& + &
b(1,1)^3 b(1,3) b(2,1) b(3,2) -b(1,1)^4 b(2,3) b(3,2)
- b(1,1)^2 b(1,2) b(1,3) b(2,1) b(3,1)
\nonumber
\\
& + & b(1,1)^3 b(1,2) b(2,3) b(3,1)
+b(1,1)^3 b(1,3) b(2,1) b(3,2) -b(1,1)^4 b(2,3) b(3,2)
\nonumber
\\
& + &
b(1,1)^2 b(1,2) b(1,3) b(2,1) b(3,1) - b(1,1)^3 b(1,3) b(2,2) b(3,1)
- b(1,1)^3 b(1,2) b(2,1) b(3,3)
\nonumber
\\
& + & b(1,1)^4 b(2,2) b(3,3) .
\ega
There are seven different terms, as
the 1st, 5th, 9th, 13th, the 2nd, 24th, the 3rd,15th, the 4th, 16th, the
6th, 10th, the 7th, 11th and the 8th, 12th terms of 16 elements of the
sum are equal. These will be collected by the Fortran part.

The output of the programs ``{\sl obk.m}'' and ``{\sl tbk.m}''
can be understood similarly:
\beq
\langle \Psi \vert T  \vert \Psi \rangle =
- t(3,3) b(1,1)^3 b(1,2) b(2,1)
+ t(3,1) b(1,1)^2 b(1,2) b(1,3) b(3,1) +  ...  ,
\eq
and
\bega
\langle \Psi \vert{\cal O}_{ij}^{(2)}
\vert \Psi \rangle & = & ({\sl W_1}+{\sl M_1}+{\sl H_1}+{\sl B_1}) v(1,1,1,1)
(b(1,2) b(1,3) b(2,1) b(3,1)
\\
& - & b(1,1) b(1,3) b(2,2) b(3,1)
 - b(1,1) b(1,2) b(2,1) b(3,3) + b(1,1)^2 b(2,2) b(3,3) ) +
\nonumber
\\
&  & ({\sl W_2}+{\sl M_2}+{\sl H_2}+{\sl B_2}) v(1,1,1,1)
(b(1,2) b(1,3) b(2,1) b(3,1)
\nonumber
\\
& - & b(1,1) b(1,3) b(2,2) b(3,1) - b(1,1) b(1,2) b(2,1) b(3,3)
+ b(1,1)^2 b(2,2) b(3,3) ) + ....
\nonumber
\ega
As the output  of  ``{\sl tbk.m}'' is too
long (560 terms), we present only the first 546
lines of  ``{\sl tbk.out}''.

The ``-1'' at the bottom of the Mathematica's output serves as
an end of file indicator for the Fortran part.

\section{Fortran part}

Due to the spin-isospin symmetry, a great number of terms of the
sums (\ref{eq:12}), (\ref{eq:13}) and (\ref{eq:14}) will be
exactly equal
in the Mathematica's output. By collecting these, the number of
terms can be reduced by about an order of magnitude.
E.g., in  the case of  $^8$He=$\alpha+n+n+n+n$ five cluster
system, the potential energy kernel contains about 50000
terms (see eq.(16)),and only 3528 of them are different.
As the Mathematica stores the symbols and operations, this collection
by Mathematica
would require enormously large memory and might be very slow.
As in the
Mathematica's output the analytical formulae
(\ref{eq:12}), (\ref{eq:13}) and (\ref{eq:14}) are represented
by integer
numbers, this simplification can efficiently be achieved by
using the Fortran language.

After collecting the identical
terms, the forms of the expressions
(\ref{eq:12}), (\ref{eq:13}) and (\ref{eq:14}) remain the same,
except
for the coefficient $c_{n}^{o},c_{n}^{ob}$ and $c_{n}^{tb}$
, where the multiplicity of each different terms appears.

In the Fortran part we can also substitute the concrete form of
the spatial part of the overlap, one-body and two-body matrix
elements of the single-particle  functions.
In this paper we limit ourselves to the case of displaced
gaussian single-particle functions of equal size parameters:
\beq
\varphi_i({\bf r}_k)=
\left( {2 \nu \over \pi} \right)^{3/4} e^{-\nu ({\bf r}-{\bf
s}_i)^2} ,
\eq
where ${\bf s}_i$ is the ``generator coordinate vector''
pointing to the centre-of-mass of the $i$th cluster (see Figure
1). For the case of nonequal size parameters or for other type
of single particle functions this part of the program can easily
be generalized. As the generator coordinates in the bra may
differ from those in the ket the vectors in the bra
and ket position will be distinguished by using a prime ($'$).

The two-body potential is represented by gaussian form factor
\beq
V(i,j)=e^{-\mu({\bf r}_i - {\bf r}_j )^2} ,
\eq
and as an example, the one-body operator is chosen to be the
kinetic energy  operator $T$.
The single particle matrix elements are \cite{Horiuchi}
\beq
b(i,j)=e^{({\bf s}_i-{\bf s'}_j)^2} ,
\label{eq:b}
\eq
\beq
t(p,q)=b(p,q) {\hbar^2 \over 2 m}
\nu (3-\nu({\bf s}_p - {\bf s'}_q)^2) ,
\label{eq:t}
\eq
\beq
v(p,q;r,t)=\theta^{3/2} b(p,r) b(q,t)
e^{-{1-\theta \over 4} \nu ({\bf s}_p -{\bf s}_q + {\bf s'}_r
-{\bf s'}_t )^2} ,
\label{eq:v}
\eq
where $\theta=\nu/(\nu+\mu)$.

By substituting the single particle matrix elements
(\ref{eq:b}), (\ref{eq:t}) and (\ref{eq:v}) into eqs.
(\ref{eq:12}), (\ref{eq:13}) and (\ref{eq:14})
the matrix elements of the wave function $\Psi$
take the form:

\par\noindent
-- overlap:
\beq
\label{eq:over}
\langle \Psi \vert \Psi \rangle =
\sum_{i=1}^{N_o} C_i^o {\rm exp}({-{1 \over 2}\nu {\underline
\bf s}^{\dagger}  A_{i}^{o}
{\underline \bf s}}) ,
\eq
\par\noindent
-- one-body:
\beq
\label{eq:over1}
\langle \Psi \vert T  \vert \Psi \rangle =
{\hbar^2 \nu \over 2 m}
\sum_{i=1}^{N_{ob}} C_i^{ob}
(3 - \nu {\underline \bf s}^{\dagger} B_{i}^{ob} {\underline \bf s})
{\rm exp}({-{1 \over 2} \nu  {\underline \bf s}^{\dagger}
A_{i}^{ob} {\underline \bf s}}) ,
\eq
\par\noindent
-- two-body:
\beq
\langle \Psi \vert{\cal O}_{ij}^{(2)}
\vert \Psi \rangle =
\theta^{3/2}\sum_{i=1}^{N_{tb}} (w C_i^{W}+ m C_i^{M}+b C_i^{B}+h C_i^{H})
{\rm exp}({-{1 \over 2}\nu {\underline \bf s}^{\dagger}
A_{i}^{tb} {\underline \bf s}
-{1-\theta \over 4} \nu (a_{i}^{tb} {\underline \bf s})^2}) ,
\eq
where $A_{i}^{o}, A_{i}^{ob}, A_{i}^{tb}$ and $B_{i}^{ob}$ are
$2N \times 2N$ matrices, $a_{i}^{tb}$ is an $2N$ dimensional
vector, and
\beq
{\underline \bf s}^{\dagger}=\left(
{\bf s}_1,
{\bf s}_2,
... ,
{\bf s}_N,
{\bf s'}_1,
{\bf s'}_2 ,
... ,
{\bf s'}_N
\right) .
\eq
After suitable angular momentum projection /cite{Brink},
these matrix elements are building blocks of the
cluster model calculations.

\subsection{Input data}

The Fortran part needs the input data stored in the file
``{\sl nuc.dat}'', and the Mathematica's output ``{\sl ok.out}'',
``{\sl obk.out}'' and ``{\sl tbk.out}''. Free format is used.

\subsection{Output data}

The Fortran programs ``{\sl ok.f}'', ``{\sl obk.f}'' and ``{\sl tbk.f}''
handle the overlap, the kinetic energy and the potential energy
kernels, respectively. The output is written in the files
``{\sl ok.inp}'', ``{\sl obk.inp}'' and ``{\sl tbk.inp}'' ordered as shown on
Table 2.

\subsection{Test run}

After collecting the identical terms, the Fortran part
substitutes the single particle matrix elements, thus e. g.
the overlap looks like:
\bega
\langle \Psi \vert \Psi \rangle & = &
2 {\rm exp}\lbrace-{ \nu \over 2}
(2({\bf s}_1-{\bf s'}_1)^2 +
({\bf s}_1-{\bf s'}_2)^2
({\bf s}_1-{\bf s'}_3)^2 +
({\bf s}_2-{\bf s'}_1)^2 +
({\bf s}_3-{\bf s'}_1)^2) \rbrace
\nonumber
\\
& - &
{\rm exp}\lbrace-{\nu \over 2}
(3({\bf s}_1-{\bf s'}_1)^2 +
({\bf s}_1 - {\bf s'}_3)^2 +
({\bf s}_2-{\bf s'}_2)^2 +
({\bf s}_3-{\bf s'}_1)^2 ) \rbrace
\nonumber
\\
& - &
{\rm exp}\lbrace-{\nu \over 2}(
-3({\bf s}_1-{\bf s'}_1)^2 +
({\bf s}_1-{\bf s'}_2)^2 +
({\bf s}_2-{\bf s'}_1)^2 +
({\bf s}_3-{\bf s'}_3)^2
) \rbrace
\nonumber
\\
& + &
{\rm exp}\lbrace-{\nu \over 2}(
+4({\bf s}_1-{\bf s'}_1)^2 +
({\bf s}_2-{\bf s'}_2)^2 +
({\bf s}_3-{\bf s'}_3)^2
) \rbrace
\nonumber
\\
& - &
{\rm exp}\lbrace-{\nu \over 2}(
3({\bf s}_1-{\bf s'}_1)^2 +
({\bf s}_1-{\bf s'}_2)^2 +
({\bf s}_2-{\bf s'}_3)^2 +
({\bf s}_3-{\bf s'}_1)^2
) \rbrace
\nonumber
\\
& - &
{\rm exp}\lbrace-{\nu \over 2}(
3({\bf s}_1-{\bf s'}_1)^2 +
({\bf s}_1-{\bf s'}_3)^2 +
({\bf s}_2-{\bf s'}_1)^2 +
({\bf s}_3-{\bf s'}_2)^2
) \rbrace
\nonumber
\\
& + &
{\rm exp}\lbrace-{\nu \over 2}(
({\bf s}_1-{\bf s'}_1)^2 +
({\bf s}_2-{\bf s'}_3)^2 +
({\bf s}_3-{\bf s'}_2)^2
) \rbrace
\nonumber
\ega
{}From this form the matrices $A^o$ in eq. (\ref{eq:over}) are
determined and written into ``{\sl ok.out}'' (see the test run output).
The output files of ``{\sl obk.out}'' and ``{\sl obk.out}'' are
too long, and only the first 53 and 57 lines are listed.

\subsection*{Acknowledgements}
The first version of this program was written while
I was Visiting Scientist at Niigata
University. I am very pleased to thank Professor Y. Suzuki
for his hospitality.
I am grateful to Dr. A. Cs\'ot\'o and Dr. A. T. Kruppa for useful
discussions.
This work was supported by the OTKA grant (No. F4348).

{}
\newpage
\begin{center}
Table 1
\end{center}
\[
\begin{array}[t]{l}
overlap \\
\alpha,\beta,n_o \\
m_1 \\
i_{1_1},j_{1_1},k_{1_1} \\
\vdots \\
i_{m_{1}},j_{m_{1}},k_{m_{1}} \\
c_{1}^o                 \\
m_2                   \\
i_{1_2},j_{1_2},k_{1_2}           \\
\vdots
\end{array}
\begin{array}[t]{c}
\ \ \ \ \ \ \ \ \ \ \ \ \ \   \\
\ \ \ \ \ \ \ \ \ \ \ \ \ \
\end{array}
\begin{array}[t]{l}
one-body \\
\alpha,\beta,n_{ob} \\
p_{1},q_{1}      \\
m_{1} \\
i_{1_1},j_{1_1},k_{1_1} \\
\vdots \\
i_{m_{1_1}},j_{m_{1_1}},k_{m_{1_1}} \\
c_{1}^{ob}                 \\
p_{2},q_{2}               \\
m_2                   \\
i_{1_2},j_{1_2},k_{1_2}           \\
\vdots
\end{array}
\begin{array}[t]{c}
\ \ \ \ \ \ \ \ \ \ \ \ \ \   \\
\ \ \ \ \ \ \ \ \ \ \ \ \ \
\end{array}
\begin{array}[t]{l}
two-body \\
\alpha,\beta \\
W_1,M_1,B_1,H_1      \\
p_1,q_1,r_1,s_1      \\
n_1                \\
m_1                  \\
i_{1_1},j_{1_1},k_{1_1} \\
\vdots \\
i_{m_{1}},j_{m_{1}},k_{m_{1}} \\
c_{1}^{tb}                 \\
m_2               \\
i_{1_1},j_{1_1},k_{1_1} \\
\vdots \\
i_{m_{2}},j_{m_{2}},k_{m_{2}} \\
c_{2}^{tb}                 \\
\vdots               \\
m_{n_1}             \\
i_{1_1},j_{1_1},k_{1_1} \\
\vdots \\
i_{m_{n_1}},j_{m_{n_1}},k_{m_{n_1}} \\
c_{n_1}^{tb}                 \\
\alpha,\beta        \\
W_2,M_2,B_2,H_2      \\
p_2,q_2,r_2,s_2      \\
n_2
m_1                  \\
\vdots
\end{array}
\]
\newpage
\begin{center}
Table 2
\end{center}
\[
\begin{array}[t]{ccc}
overlap & & \\
N_o     & & \\
C^o_1   & & \\
\left( A^o_1 \right)_{11} &  . . .   & \left( A^o_1 \right)_{1N} \\
\vdots & & \vdots \\
\left( A^o_1 \right)_{N1} &  . . .   & \left( A^o_1 \right)_{NN} \\
C^o_2   & & \\
\vdots & &
\end{array}
\begin{array}[t]{c}
\ \ \ \ \ \ \ \ \ \   \\
\ \ \ \ \ \ \ \ \ \
\end{array}
\begin{array}[t]{ccc}
one-body & & \\
N_{ob}     & & \\
C^{ob}_1   & & \\
\left( A^{ob}_1 \right)_{11} &  . . .   & \left( A^{ob}_1 \right)_{1N} \\
\vdots & & \vdots \\
\left( A^{ob}_1 \right)_{N1} &  . . .   & \left( A^{ob}_1 \right)_{NN} \\
\left( B^{ob}_1 \right)_{11} &  . . .   & \left( B^{ob}_1 \right)_{1N} \\
\vdots & & \vdots \\
\left( B^{ob}_1 \right)_{N1} &  . . .   & \left( B^{ob}_1 \right)_{NN} \\
C^{ob}_2   & & \\
\vdots & &
\end{array}
\begin{array}[t]{c}
\ \ \ \ \ \ \ \ \ \ \\
\ \ \ \ \ \ \ \ \ \
\end{array}
\begin{array}[t]{ccc}
two-body & & \\
N_{tb}     & & \\
C^{W}_1,C^{M}_1,C^{B}_1,C^{H}_1  & & \\
\left( A^{tb}_1 \right)_{11} &  . . .   & \left( A^{tb}_1 \right)_{1N} \\
\vdots & & \vdots \\
\left( A^{tb}_1 \right)_{N1} &  . . .   & \left( A^{tb}_1 \right)_{NN} \\
\left( a^{tb}_1 \right)_{1} &  . . .   & \left( a^{tb}_1 \right)_{N} \\
C^{W}_2,C^{M}_2,C^{B}_2,C^{H}_2  & & \\
\vdots & &
\end{array}
\]

\end{document}